\documentclass[aps,prl,twocolumn,amsmath,amssymb,nofootinbib,superscriptaddress,floatfix,reprint,longbibliography]{revtex4-1}

\usepackage[dvips]{graphicx}
\usepackage{latexsym}
\usepackage{amsmath}
\usepackage{amsfonts}
\usepackage{amssymb}
\usepackage{bm}
\usepackage{color}
\usepackage{txfonts}
\usepackage{float}
\usepackage{url}
\usepackage[colorlinks=true, urlcolor=blue, linkcolor=blue, citecolor=blue, pdftex]{hyperref}
\usepackage{ulem}
\usepackage{graphicx}
\usepackage{extpfeil}
\usepackage{subfigure}
\usepackage{physics}
\usepackage{setspace}
\DeclareUnicodeCharacter{2212}{-}

\begin{document}

	\newcommand{\fig}[2]{\includegraphics[width=#1]{#2}}
	\newcommand{\la}{{\langle}}
	\newcommand{\ra}{{\rangle}}
	\newcommand{\dg}{{\dagger}}
	\newcommand{\upa}{{\uparrow}}
	\newcommand{\dna}{{\downarrow}}
	\newcommand{\ab}{{\alpha\beta}}
	\newcommand{\ias}{{i\alpha\sigma}}
	\newcommand{\ibs}{{i\beta\sigma}}
	\newcommand{\hH}{\hat{H}}
	\newcommand{\hn}{\hat{n}}
	\newcommand{\hc}{{\hat{\chi}}}
	\newcommand{\hU}{{\hat{U}}}
	\newcommand{\hV}{{\hat{V}}}
	\newcommand{\br}{{\bf r}}
	\newcommand{\bk}{{{\bf k}}}
	\newcommand{\bq}{{{\bf q}}}
	\def\gsim{~\rlap{$>$}{\lower 1.0ex\hbox{$\sim$}}}
	\setlength{\unitlength}{1mm}
	\newcommand{{\vhf}}{$\chi^\text{v}_f$}
	\newcommand{{\vhd}}{$\chi^\text{v}_d$}
	\newcommand{{\vpd}}{$\Delta^\text{v}_d$}
	\newcommand{{\ved}}{$\epsilon^\text{v}_d$}
	\newcommand{{\vved}}{$\varepsilon^\text{v}_d$}
	\newcommand{\pprl}{Phys. Rev. Lett. \ }
	\newcommand{\pprb}{Phys. Rev. {B}}
    \newcommand{\LNO}{La$_3$Ni$_2$O$_7$ }

 \title{Self-doped Molecular Mott Insulator for Bilayer High-Temperature Superconducting \LNO}
\author{Zhan Wang}
\thanks{These authors contributed equally.}
\affiliation{Beijing National Laboratory for Condensed Matter Physics and Institute of Physics, Chinese Academy of Sciences, Beijing 100190, China}
\affiliation{Kavli Institute for Theoretical Sciences, University of Chinese Academy of Sciences, Beijing, 100190, China}

\author{Heng-Jia Zhang}
\thanks{These authors contributed equally.}
\affiliation{Kavli Institute for Theoretical Sciences, University of Chinese Academy of Sciences, Beijing, 100190, China}
    
\author{Kun Jiang}
\email{jiangkun@iphy.ac.cn}
\affiliation{Beijing National Laboratory for Condensed Matter Physics and Institute of Physics, Chinese Academy of Sciences, Beijing 100190, China}
\affiliation{School of Physical Sciences, University of Chinese Academy of Sciences, Beijing 100190, China}

\author{Fu-Chun Zhang}
\email{fuchun@ucas.ac.cn}
\affiliation{Kavli Institute for Theoretical Sciences, University of Chinese Academy of Sciences,
	Beijing, 100190, China}

\date{\today}

\begin{abstract}
   The bilayer structure of recently discovered high-temperature superconducting nickelates La$_3$Ni$_2$O$_7$ provides a new platform for investigating correlation and superconductivity. Starting from a bilayer Hubbard model, we show that there is a molecular Mott insulator limit formed by the bonding band owing to Hubbard interaction $U$ and large interlayer coupling. This molecular Mott insulator becomes self-doped due to electrons transferred to the antibonding bands at a weaker interlayer coupling strength. The self-doped molecular Mott insulator is similar to the doped Mott insulator studied in cuprates. We propose La$_3$Ni$_2$O$_7$ to be a self-doped molecular Mott insulator, whose molecular Mott limit is formed by two nearly degenerate antisymmetric $d_{x^2-y^2}$ and $d_{z^2}$ orbitals. Partial occupation of higher energy symmetric $d_{x^2-y^2}$ orbital leads to self-doping, which may be responsible for high-temperature superconductivity in La$_3$Ni$_2$O$_7$. The effects of Hund's coupling $J_H$ on the low-energy spectra are also studied via exact diagonalization. The proposed low-energy theory for La$_3$Ni$_2$O$_7$ is found to be valid in a wide range of $U$ and $J_H$.
\end{abstract}

\maketitle

\section{Introduction}
The discovery of high-temperature superconductivity (SC) in cuprates greatly challenges our understanding of condensed matter physics~\cite{keimer_review}. The underlying physics of cuprates is widely believed to be deeply related to electron correlation and their parent Mott insulator states~\cite{doping_mott,Anderson_2004}. Hence, doping a Mott insulator is one central role for realizing a high-temperature superconductor. Following this idea, it was proposed that doping the Mott nickelates could also lead to high-temperature superconductivity~\cite{rice_PhysRevB.59.7901}. This idea became realized in thin films of the ``infinite-layer" nickelates (Sr,Nd)NiO$_2$ in 2019~\cite{lidanfeng,lidanfeng3,lidanfeng2}, which opened the Nickel age of superconductivity ~\cite{norman}.
Recently, a new type of nickelates La$_3$Ni$_2$O$_7$ (LNO) was successfully synthesized with a high-temperature superconducting transition $T_c\sim80$K under high pressure~\cite{meng_wang,meng_wang2,chengjg,chengjg_poly,yuanhq,chengjg_crystal}.
Many theoretical proposals have been developed to uncover its superconducting mechanism~\cite{yaodx,wangqh,zhanggm,dagotto1,werner,yangyf2,Lechermann_PhysRevB.108.L201121,yangf,sugang,kun_cpl,Kuroki,Luzy_PhysRevB.109.115114,dagotto2,wucj,taoxiang,ryee2023critical,guyh}.

Compared to the essential CuO$_2$ layer in cuprates, the central ingredient for La$_3$Ni$_2$O$_7$ is the bilayer NiO structure.
The valence charge of Ni is 3d$^{7.5+}$, seemingly far away from any strong correlation limit. It is then interesting and important what is the Mott limit of the bilayer system. Theoretically, the bilayer systems have been explored many years ago, especially the bilayer Hubbard model~\cite{Scalapino_PhysRevB.45.5577,Hanke_PhysRevB.50.4159,Scalettar_PhysRevB.50.13419,Santos_PhysRevB.51.15540,okamoto_PhysRevB.75.193103,scalettar_PhysRevB.77.144527,fabrizio_PhysRevB.80.224524,Maier_PhysRevB.84.180513}. In this work, we wish to show that the strong correlation limit of this bilayer structure is described by the two-orbital molecular Mott insulator, and self-doping this molecular Mott insulator leads to the high-temperature superconductivity.

\begin{figure}
    \centering
    \includegraphics[width=\columnwidth]{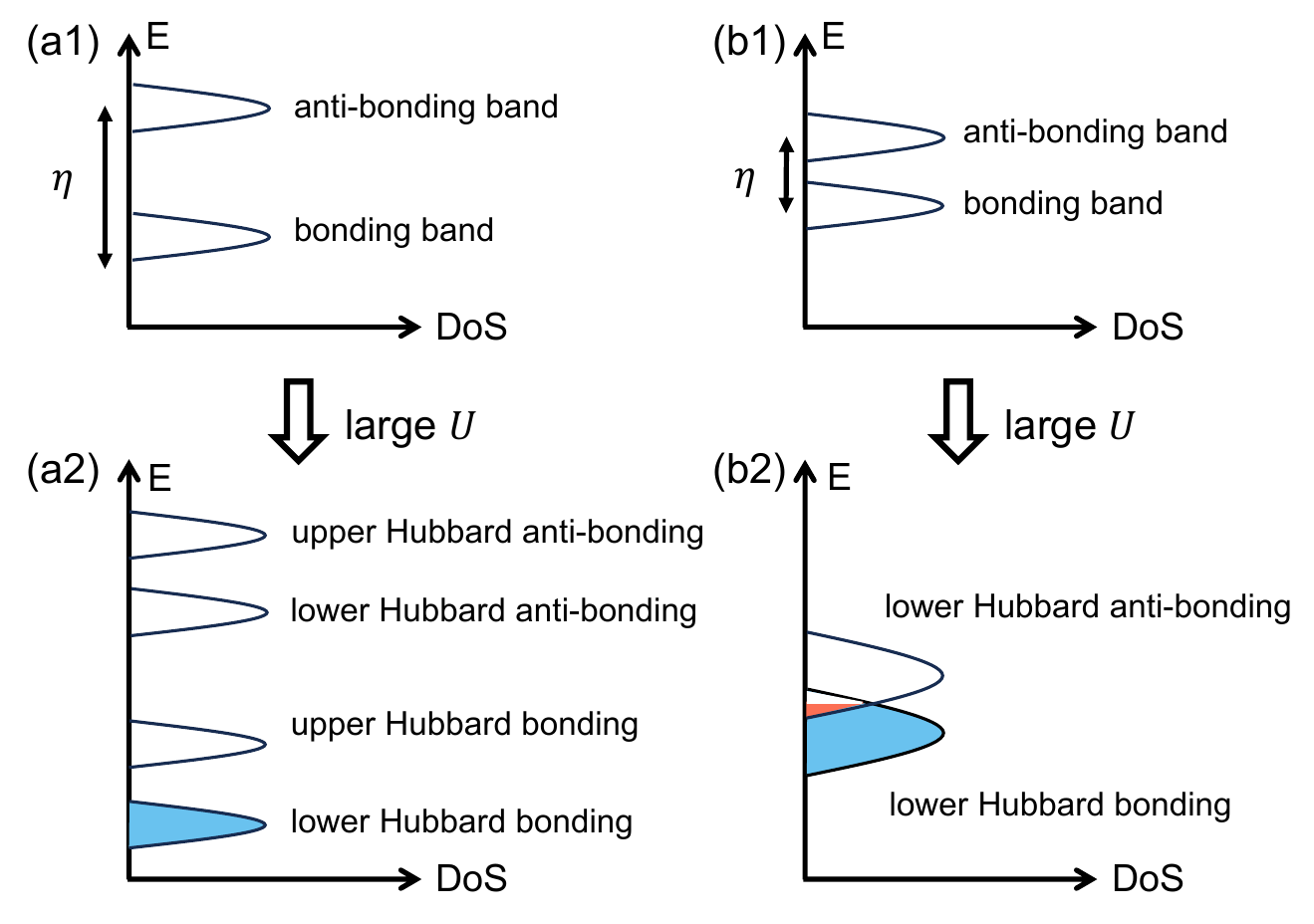}
    \caption{Self-doped Molecular Mott insulator. (a1) and (b1) show bands formed by molecular bonding and anti-bonding orbitals that are separated by different values of the molecular orbital splitting $\eta$. (a2) and (b2) schematically show the upper and lower Hubbard band resulting from onsite repulsion $U$ in different limits of $\eta/U$. The filling considered here is one electron per molecule. For $\eta\gg U$ as shown in (a2), the four Hubbard bands are isolated from each other and the electrons will fill the lower bonding Hubbard band only. This corresponds to the scenario of a molecular Mott insulator. In the scenario of (b2) where $\eta< U$, overlap between the lower Hubbard bands of the two molecular orbitals is found due to small $\eta$. The two upper Hubbard bands at higher energy are empty and are not plotted here. At the quarter filling, the electrons now resides primarily in the lower Hubbard bonding band, with a small portion in the lower anti-bonding Hubbard band, giving rise to the self-doped molecular Mott insulator.}
    \label{schematic}
\end{figure}
 
\section{Self-doped Molecular Mott Insulator}
To introduce the concept of the self-doped molecular Mott insulator, we start by considering two identical layers of single-band electrons stacked together to form a bilayer system with interlayer hopping parameter  denoted by $\frac{\eta}2$. In the non-interacting limit, the interlayer hopping separates the electron dispersion into a bonding and an anti-bonding band as illustrated in Fig.~\ref{schematic} (a1, b1). The Wannier orbitals for the bonding and the anti-bonding bands are linear combinations of atomic orbitals belonging to different layers. Hence, these Wannier orbitals should be considered as \textit{molecular orbitals}.

Next, taking Hubbard interaction $U$ into account, the electron spectrum splits into the upper and the lower Hubbard bands. The interlayer hopping $\frac{\eta}2$ competes with the onsite repulsion $U$ and the bandwidth, leading to different insulating limits beyond the upper and lower Hubbard bands in the single-band Hubbard model. There are two typical scenarios as schematically illustrated in Fig.~\ref{schematic}. In the first case where $\eta$ is large, as depicted in Fig.~\ref{schematic} (a2), the four Hubbard bands are well separated from each other. At the quarter filling, that is one electron per molecule, we get the singly occupied lower Hubbard bonding band. The effective theory, in terms of molecular orbitals, shares the same feature of the single-band Mott insulator. Therefore, it is a \textit{molecular Mott insulator}.
An intriguing scenario is expected when the molecular energy splitting $\eta$ is not too large, as shown in Fig.~\ref{schematic} (b2). In this case, the two lower Hubbard bands of the bonding and the anti-bonding orbitals overlap with each other. Consequently, a small number of electrons can occupy the anti-bonding Hubbard band. Consequently, the bonding Hubbard band becomes hole-doped even at the quarter filling. We term this scenario as the \emph{self-doped molecular Mott insulator}.

\begin{figure}
    \centering
    \includegraphics[width=\columnwidth]{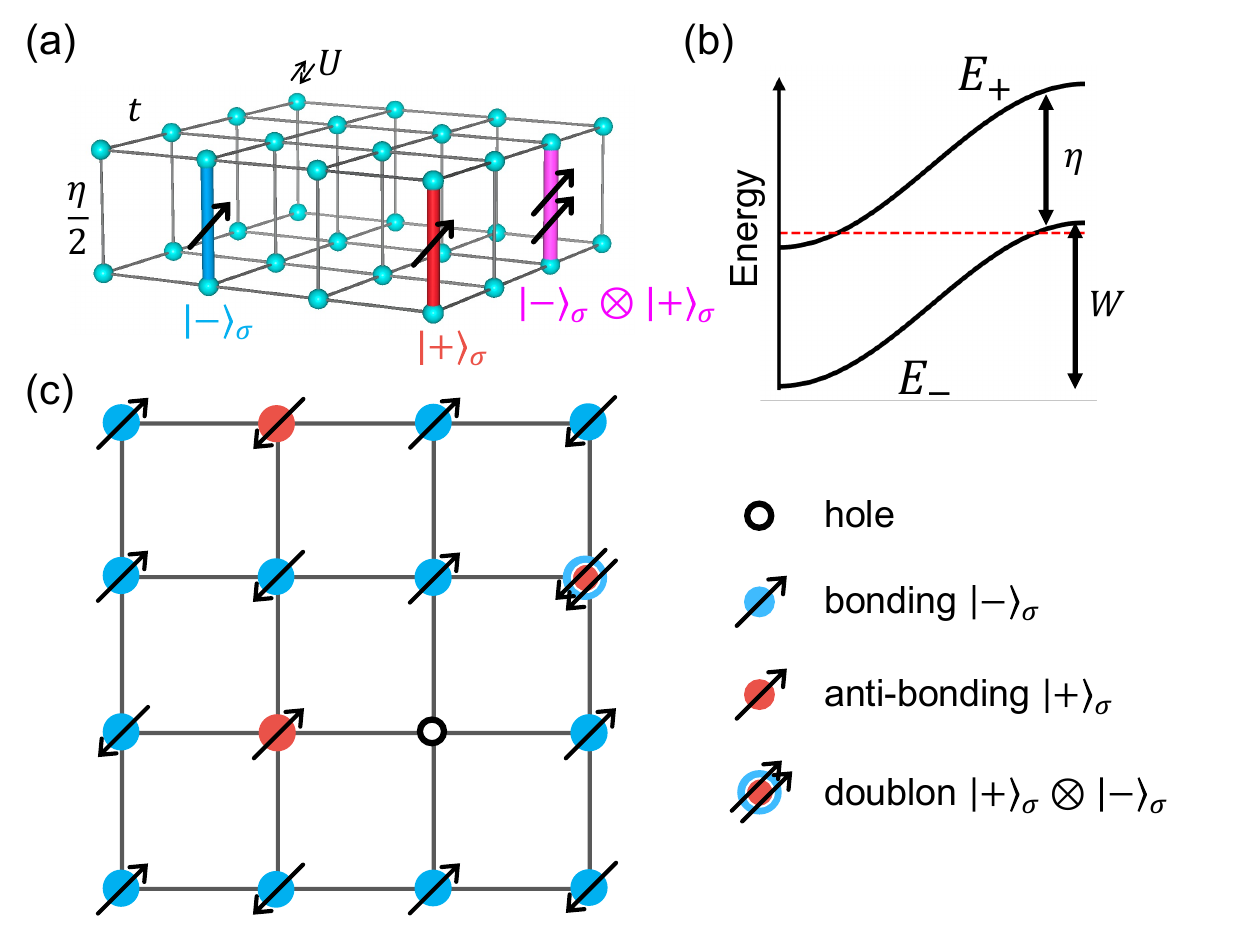}
    \caption{Self-doped molecular Mott insulator from the bilayer Hubbard model. (a) schematic illustration of the bilayer Hubbard model with interlayer hopping is denoted by $\frac{\eta}{2}$, the intra-layer hopping $t$ and onsite repulsion $U$. The bonding and anti-bonding orbitals in the molecular basis is denoted by blue and red, respectively. The doublon state is denoted by purple. (b) Schematic energy dispersion of the bonding and the anti-bonding bands. At $\eta<W$, the ground state at the filling of one electron per molecule contains both the bonding and anti-bonding orbitals. (c) Schematic illustration of the self-doped molecular Mott insulator with large U. The blue and red circles denote the electrons in the bonding and the anti-bonding orbitals, respectively. Note the depicted doublon state with both orbitals occupied doesn’t cost U and is also allowed in the low energy range, see eq.~\eqref{doublon} in the text.}
    \label{bilayer-Hubbard}
\end{figure}

As a concrete model to realize the self-doped molecular Mott insulator, consider the bilayer square lattice Hubbard model depicted in Fig.~\ref{bilayer-Hubbard} (a). The Hamiltonian can be written as:
\begin{equation}
    H_{\text{bilayer}}=H_t+H_U.
\end{equation}
with the kinetic part $H_t$ and the interaction part $H_U$ given by:
\begin{align}
    H_t&=-t\sum_{\langle ij\rangle\sigma}\sum_{l=t,b}c_{il\sigma}^\dagger c_{jl\sigma}+\frac{\eta}{2}\sum_{i,\sigma} c_{it\sigma}^\dagger c_{ib\sigma}+\text{h.c.}\\
    H_U&=U\sum_{i}\sum_{l=t,b} n_{il\uparrow}n_{il\downarrow}.
\end{align}
Here the layer index $l=t,b$ denotes the top and bottom layer, respectively. $c_{il\sigma}^\dagger$ creates an electron with spin $\sigma$ at planar site $i$ of layer $l$ and $n_{il\sigma}$ is the corresponding electron number operator. The two layers share the same in-plane nearest neighbor hopping amplitude $t$ and the repulsive Hubbard interaction $U$. The interlayer hopping is $\frac{\eta}{2}$ between the corresponding sites from the two layers. We assume $\eta>0$.

The kinetic part $H_t$ can be diagonalized under the basis of the interlayer molecular bonding (antisymmetric, $-$) and anti-bonding (symmetric, $+$) orbitals, namely $c_{i\pm\sigma}=(c_{it\sigma}\pm c_{ib\sigma})/\sqrt{2}$ as labeled by blue and red in Fig.~\ref{bilayer-Hubbard} (a). The obtained molecular energy dispersions are $E_{k\pm}=\epsilon_k\pm \frac{\eta}{2}$ with $\epsilon_k=-2t[\cos k_x+\cos k_y]$. The molecular energy splitting between the bonding and the anti-bonding bands is $\eta$. The bonding band ${E_{k-}}$ has lower energy, as schematically depicted in Fig.~\ref{bilayer-Hubbard} (b). The band width $W$ is determined by the in-plane hopping and the band energy splitting corresponds to $\eta$. For $\eta\lesssim W$, there is a finite overlap between the bonding and the anti-bonding bands as highlighted by red dashed line in Fig.~\ref{bilayer-Hubbard} (b).

At large Hubbard $U$, the electron band further splits into the upper and lower Hubbard band. If $\eta\lesssim W\ll U$, we expect that the local electrons are mostly found in the lower bonding Hubbard band $E_-$, with a small number of electrons in the anti-bonding Hubbard band $E_+$, realizing the self-doped molecular Mott insulator through the scenario similar to the one depicted in Fig.~\ref{schematic} (b).

It is well known that the doped Mott insulator can be well described by the effective t-J model~\cite{doping_mott,Anderson_2004,zhang_rice_PhysRevB.37.3759}.
For this self-doped molecular Mott insulator, we can also write down an effective $t-J$ model. 
Compared to the single band $t-J$ model, there are four possible energy-allowed states: hole, singly-occupied anti-bonding $|+\rangle_\sigma$, singly-occupied bonding $|-\rangle_\sigma$ and a special doublon state $|+\rangle_\sigma \otimes |-\rangle_\sigma$, as summarized in Fig.~\ref{bilayer-Hubbard} (c). Note this doublon state is formed by two electrons of the same spin occupying both the bonding and the anti-bonding orbitals. This particular configuration is found to be immune from double onsite occupancy constraints as it doesn't cost $U$. One can see this through its wavefunction:
\begin{align}
    c_{i+\sigma}^\dagger\otimes c_{i-\sigma}^\dagger&=\frac1{\sqrt{2}}(c_{it\sigma}^\dagger+c_{ib\sigma}^\dagger)\otimes\frac1{\sqrt{2}}(c_{it\sigma}^\dagger-c_{ib\sigma}^\dagger)\notag\\
    &=c_{it\sigma}^\dagger \otimes c_{ib\sigma}^\dagger.
    \label{doublon}
\end{align}
In this state, one electron ins on the top and the other electron on the bottom layer, which does not cost any $U$. At quarter filling, the total electron number equals one per molecule. The formation of a doublon will generate a hole at another molecular site, as shown in Fig.~\ref{bilayer-Hubbard} (c), leading to the picture of the self-doping effect in real space.

Therefore, the effective $t-J$ model can be written down as, in addition  to an weakly correlated band for $|+\rangle_\sigma$ electrons,
\begin{align}
    H_{t-J}=&-t^- \sum_{\langle ij\rangle,\sigma} \hat{P}_G (c_{i-\sigma}^\dagger c_{j-\sigma}+\text{h.c.})\hat{P}_G +J\sum_{\langle ij\rangle}\bm{S}_{i-}\cdot\bm{S}_{j-}
\end{align}
Here $t^-$ is the bare hopping amplitude associated with the antisymmetric bonding electrons and $\bm{S}_{i-}=\frac12c_{i-s_1}^\dagger \bm{\sigma}_{s_1s_2} c_{i-s_2}$ denotes the local spin operator of the bonding electron with $s_1,s_2$ representing the spin-$1/2$ indices. $\hat{P}_G$ is the projection operator to the above energy-allowed configurations in the large-$U$ limit. Because the correlation effect in anti-bonding orbitals is weak, we only treat them as self-doping to the bonding $t-J$ model. Owing to the self-doping effect, the $t-J$ model can naturally induce a $d$-wave superconductivity, as widely studied in cuprates~\cite{Anderson_2004,doping_mott,zhang-d-wave}.

\section{Effective Model for $\text{La}_3\text{Ni}_2\text{O}_7$}
Next we turn to the effective model for La$_3$Ni$_2$O$_7$. Through the local electronic structure, we demonstrate that the low energy physics of the \LNO can be described by a two-orbital self-doped molecular Mott insulator. 

\begin{figure}
    \centering
    \includegraphics[width=\columnwidth]{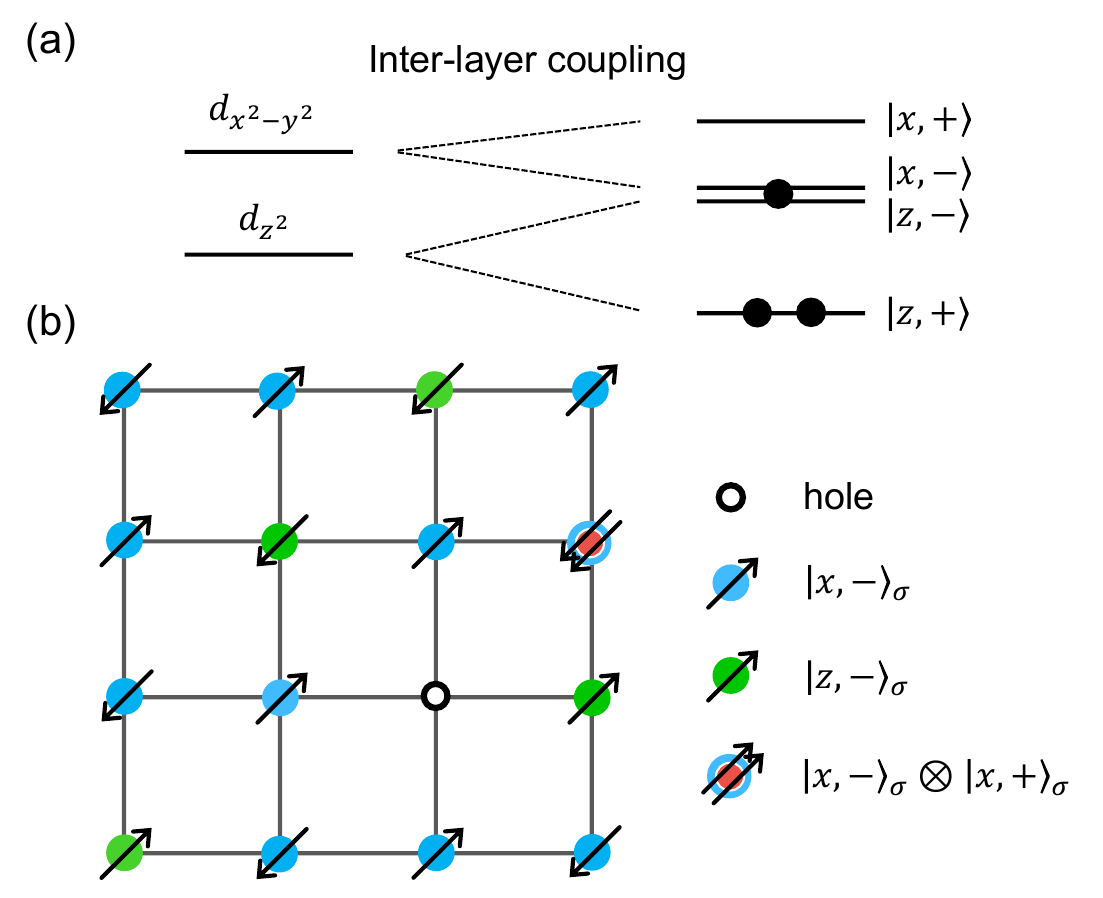}
    \caption{Model for \LNO: two-orbital self-doped Molecular Mott insulator. (a) Local electronic orbitals of the interlayer pair $(\text{Ni}_2)^{5+}$. For each Ni atom, $3d$ $t_{2g}$ are fully filled and not plotted here. With interlayer coupling $t_\perp^{x,z}$, the two sets of $e_g$ orbitals further split into four molecular orbitals. Two electrons occupy the $|z,+\rangle$ orbital which is much lower in energy than the other three orbitals. The remaining one electron predominantly occupies one of the two anti-bonding orbitals $|x,-\rangle_\sigma$ and $|z,-\rangle_\sigma$, with a small portion in the bonding orbital $|x,+\rangle_\sigma$, corresponding to the $\alpha$-band in the DFT calculation~\cite{yaodx}. (b) Schematic illustration of the self-doped Molecular Mott insulator. In the Mott limit, electrons can only occupy either one of the $|x,-\rangle_\sigma$ and $|z,-\rangle_\sigma$ orbitals due to the strong onsite Hubbard repulsion, as depicted in blue and green, respectively. A small amount of doublon states, namely $|x,-\rangle_\sigma \otimes|x,+\rangle_\sigma$ as depicted in blue-red circles, can exist due to no cost in U. With the total number of electrons fixed, formation of a doublon will generate a hole, leading to the self-doped molecular Mott insulator.}
    \label{mmi-327}
\end{figure}

The crystal structure of the \LNO material under high pressure has the space group $Fmmm$, with stacked bilayer NiO$_2$ planes formed by corner-sharing Ni-O octahedra along the apical direction. Due to the octahedra crystal field and Jahn-Teller distortion, the partially-filled $e_g$ orbitals form the atomic occupancy of bilayer Ni$^{2.5+}$ atoms, as illustrated in Fig.~\ref{mmi-327} (a). Taking the interlayer coupling into account, the molecular orbitals from the $e_g$ orbitals further split into four energy levels in Fig.~\ref{mmi-327} (a). 

Owing to the different wavefunction symmetry of the $d_{z^2}$ and the $d_{x^2-y^2}$ orbital along the $z$-direction, the interlayer hoppings $t_\perp^{x,z}$ are different. For $d_{z^2}$ orbitals, the interlayer hopping is mediated by the $p_z$ orbital of the apical oxygen ions between the top (t) and bottom (b) layer. The relevant hopping path is $d_{z^2,t}\rightarrow p_z\rightarrow d_{z^2,b}$. The wave function symmetry of the O-$p_z$ orbital yields a sign change, resulting in $t^z_\perp<0$. While for $d_{x^2-y^2}$ orbitals, the same hopping path $d_{x^2-y^2,t}\rightarrow p_z\rightarrow d_{x^2-y^2,b}$ gives zero amplitude and the interlayer hopping is mediated by other paths including more sites. One can therefore expect the interlayer hopping $|t_{\perp}^x|<|t_{\perp}^z|$. More concrete values of the interlayer hoppings require detailed calculations and we take $t_\perp^x>0$ as obtained from the density functional theory (DFT) results~\cite{2025Jiang}.

In \LNO, the low energy physics is described by 3 electrons per Ni$_2$ molecular ion, filling the two sets of $e_g$ orbitals. The filling is depicted in Fig.~\ref{mmi-327} (a).  The molecular bonding orbital of the $d_{z^2}$ is a symmetric combination of the interlayer states, denoted as $|z,+\rangle$, which has much lower energy and is fully occupied. We therefore drop this orbital in the following discussions in this section. It is interesting that the energy splitting due to Jahn-Teller distortions is comparable to the $t_{\perp}^{z}$ in \LNO as reported in~\cite{YXWang_PhysRevB.110.205122}. This leads to the nearly degenerate antisymmetric $e_g$ orbitals denoted as $|x,-\rangle$ and $|z,-\rangle$. Below we consider the symmetric $|x,+\rangle$ orbital to be slightly higher in energy than $|x,-\rangle$, and show that the one-electron-filled antisymmetric $e_g$ orbitals form a molecular Mott insulator and self-doping this molecular Mott insulator owing to $|x,+\rangle$ gives us the physics of \LNO.

The Hamiltonian of \LNO contains 3 parts: $H=H_S+H_A+H_{SA}$. For the symmetric orbital $|x,+\rangle_\sigma$, the Hamiltonian $H_S$ reads:
\begin{equation}
    H_S=-\sum_{\langle ij\rangle,\sigma}t_{ij}^+(c_{i,x+,\sigma}^\dagger c_{j,x+,\sigma}+\text{h.c.})+\epsilon_{x+}\sum_{i\sigma}c_{i,x+,\sigma}^\dagger c_{i,x+,\sigma}, 
    \label{H1}
\end{equation}
where $c_{i,x+,\sigma}^\dagger$ is the creation operator of $|x,+\rangle_\sigma$ with spin $\sigma$, $t_{ij}^+$ denotes the in-plane hopping amplitude and $\epsilon_{x+}>0$ denotes the molecular energy shift due to $t_\perp^x$. 

$H_A$ is for antisymmetric orbitals $|x,-\rangle_\sigma$ and $|z,-\rangle_\sigma$, which we assume to be degenerate for simplicity,
\begin{align}
    H_A=&-\sum_{\langle ij\rangle,\alpha\alpha^\prime,\sigma}t_{ij}^{\alpha\alpha^\prime} (c_{i,\alpha-,\sigma}^\dagger c_{j,\alpha^\prime-,\sigma}+\text{h.c.})+\sum_{i\alpha}U_A^{\alpha\alpha} n_{i,\alpha-,\uparrow} n_{i,\alpha-,\downarrow}\notag\\
    &+ \sum_{i} U_A^{xz} (n_{i,x-,\uparrow}+n_{i,x-,\downarrow})(n_{i,z-,\uparrow}+n_{i,z-,\downarrow}).
\end{align}
Here $c_{i,\alpha-,\sigma}^\dagger$ with $\alpha=x,z$ is the creation operator associated with $|x,-\rangle_\sigma$ and $|z,-\rangle_\sigma$ and $n_{i,\alpha-,\sigma}$ is the electron number operator. The in-plane hopping amplitudes $t_{ij}^{\alpha\alpha^\prime}$ corresponds to the electrons hopping between the nearest neighbor anti-symmetric molecular orbitals $|\alpha,-\rangle$ and $|\alpha^\prime,-\rangle$. The Hubbard interactions between electrons in the antisymmetric molecular orbitals $|\alpha,-\rangle$ and $|\alpha^\prime,-\rangle$ are denoted by $U_A^{\alpha\alpha^\prime}$, which is related to the onsite repulsion $U$ for the atomic orbitals. We ignore the Hund's coupling term and will come back to its effects in the next section. Note that $H_{SA}$ denotes the coupling between the symmetric and anti-symmetric orbitals, in which the pair hopping term vanishes due to the symmetry.

In the large $U$ limit, the Hamiltonian $H_A$ alone describes a molecular Mott insulator with two nearly degenerate orbitals at one-electron filling. This is further justified through solving the local electronic structure of the two sets of $e_g$ orbitals in the large-$U$ limit, as provided in the supplementary material. This $e_g$-orbital degenerate Mott insulator has been discussed with equal importance of spin and orbital degrees of freedom~\cite{kk_model,1978Castellani,Khomskii_2014}. In \LNO, the molecular energy $\epsilon_{x+}$ is small and the introduction of a small number of $|x,+\rangle_\sigma$ will cost molecular energy but gain kinetic energy. Similar to the second scenario mentioned in the single orbital bilayer Hubbard model in Fig.~\ref{schematic} (b2), a small number of $|x,+\rangle_\sigma$ will introduce the self-doping effect which generates the doublon state $|x,-\rangle_\sigma\otimes|x,+\rangle_\sigma$ together with the self-doped holes, as depicted in Fig.~\ref{mmi-327} (b). Therefore, we propose that the effective theory of the \LNO system is described by the two-orbital self-doped molecular Mott insulator.

The low-energy effective Hamiltonian in the large-$U$ limit can be derived following the work by Castellani et al~\cite{1978Castellani}. Because the bonding orbitals $|x,+\rangle_\sigma$ are weakly correlated, its effects can be well approximated by doping additional holes into the molecular Mott insulator described by $H_A$. At $J_H=0$, the Hubbard interaction associated with the antisymmetric molecular orbitals $U_A^{xz}=U_A^{xx}=U_A^{zz}=U_A$. We define spin operator $\bm{S}_{i\alpha\beta}=\frac12 c_{i,\alpha-}^\dagger \bm{\sigma} c_{i,\beta-}$ and the density operator $n_{i\alpha\beta}=c_{i,\alpha-}^\dagger c_{i,\beta-}$, with $c_{i,\beta-}=(c_{i,\beta-,\uparrow}, \, c_{i,\beta-,\downarrow})^T$. Note that in the case with $\alpha=\beta$, $\bm{S}_{i\alpha\alpha}$ and $n_{i\alpha\alpha}$ are just the spin and density operator associated with the electrons in the same anti-symmetric orbital $\alpha-$, respectively. The effective interaction Hamiltonian derived from $H_A$ can be written as:
\begin{widetext}
\begin{align}
    H_{A,int}=&\frac{4}{U_A}\sum_{\langle ij\rangle}\sum_{\alpha\alpha^\prime\beta\beta^\prime}
\left[t_{ij}^{\alpha\beta}t_{ij}^{\alpha^\prime\beta^\prime}\bm{S}_{i\alpha\alpha^\prime} \cdot \bm{S}_{j\beta^\prime\beta} - \frac14 \Big(t_{ij}^{\alpha\beta}t_{ij}^{\alpha^\prime\beta^\prime} -   t_{ij}^{\alpha\bar{\beta^\prime}}t_{ij}^{\alpha^\prime\bar{\beta}}(-1)^{\delta_{\beta\beta^\prime}} - t_{ij}^{\bar{\alpha}\beta^\prime}t_{ij}^{\bar{\alpha^\prime}\beta }(-1)^{\delta_{\alpha\alpha^\prime}} \Big)n_{i\alpha\alpha^\prime} n_{j \beta^\prime\beta}\right].
    \label{Hint}
\end{align}
\end{widetext}
Here the orbital index can take either $x$ or $z$, referring to the two antisymmetric orbitals $|x,-\rangle$ and $|z,-\rangle$. The index with bar such as $\bar{\beta}$ refers to the other orbital. This Hamiltonian is equivalent to Kugel-Khomskii (KK) Hamiltonian~\cite{kk_model} in the limit $J_H=0$. 

In the self-doping regime, the holes are introduced by electrons transferring to the $|x,+\rangle$ orbital, resulting in the hole density $n_h=n_{x+}$. More holes can be introduced through external chemical doping so that $n_h>n_{x+}$. To provide a physical picture, we focus on the self-doping regime in the following discussions. It is essential to elucidate the correlation effect of different types of particles appropriate to the \LNO system. With one electron per site, the $|x,-\rangle_\sigma$ and $|z,-\rangle_\sigma$ electrons can only move to neighboring empty sites, which is of the density $\sim n_{x+}/2<<1$. Therefore, electrons occupying the two anti-symmetric orbitals are highly correlated. The symmetric electrons $|x,+\rangle_\sigma$, on the other hand, can hop to the neighboring anti-symmetric $|x,-\rangle_\sigma$ orbital of the same spin, with density $\sim n_{x-}/2\sim1/2$, hence are weakly correlated. We expect the superconductivity is primarily induced by the $|x,-\rangle_\sigma$ and $|z,-\rangle_\sigma$ electrons.

The in-plane hopping of the two anti-symmetric orbitals are strongly renormalized due to the correlation effect. The hopping part in the large-$U$ limit follows from the first term in $H_A$:
\begin{equation}
    H_{A,t}=-\sum_{\langle ij\rangle,\alpha\alpha^\prime}t_{ij}^{\alpha\alpha^\prime} \hat{P}_G(c_{i,\alpha-,\sigma}^\dagger c_{j,\alpha^\prime-,\sigma}+\text{h.c.})\hat{P}_G.
    \label{H0}
\end{equation}
Here $\hat{P}_G$ denotes the projection onto the low-energy Hilbert space in the large-$U$ limit. The molecular hopping parameters $t_{ij}^{\alpha\alpha^\prime}$ can be obtained from the first-principle calculations~\cite{dagotto2,2025Jiang}.

The superconducting pairing symmetry is analyzed using the renormalized mean field theory~\cite{zhang-d-wave}. The effective Hamiltonian $H_{\text{eff}}$ of the self-doped molecular Mott insulator is given by:
\begin{eqnarray}
    H_{\text{eff}}=H_{A,t}+H_{A,int},
    \label{model}
\end{eqnarray}
where we set $t_{ij}^+$ in $H_S$ equals $t_{ij}^{xx}$ in $H_{A,t}$. This is a two-orbital $t-J$ model, where $H_{A,t}$ is the kinetic energy term, describing a renormalized band of anti-symmetric orbitals whose Fermi surface is similar to that given by DFT calculations ($\beta$-band in~\cite{2025Jiang}). In the self-doping regime, the number of self-doped holes $n_h$ is controlled by $\epsilon_{x+}$ in $H_S$ with $n_h=n_{x+}$. To account for the double occupancy constraint imposed by $\hat{P}_G$ in $H_{A,t}$, we introduce the Gutzwiller factor $g_{t}^{\alpha\alpha^\prime}=\frac{2n_h}{\sqrt{(2-n_{\alpha-})(2-n_{\alpha^\prime-})}}$~\cite{slave-boson,gw-PhysRevB.76.155102}, where $n_h$ denotes the number of self-doped holes, and $n_{\alpha-}$ with $\alpha=x,z$ is the total electron number associated with the molecular orbital.

The pairing mean field in the singlet channel is defined as:
\begin{equation}
    \Delta_{ij}^{\alpha\alpha^\prime}=\langle c_{i,\alpha-,\uparrow}^\dagger c_{j,\alpha^\prime-,\downarrow}^\dagger\rangle-\langle c_{i,\alpha-,\downarrow}^\dagger c_{j,\alpha^\prime-,\uparrow}^\dagger\rangle.
    \label{pairing}
\end{equation}
with $\alpha,\alpha^\prime\in\{x,z\}$. The pairing therefore contains contributions from both the inter- and intra-orbital processes. The mean field Hamiltonian and the self-consistent equations are derived in the supplementary material. In comparison to the single orbital $t-J$ model, there are two types of nearest neighbor pairing order parameters: the intra-orbital pairing $\Delta^{xx},\,\Delta^{zz}$ and the interorbital pairing $\Delta^{xz}$, and their competition leads to different pairing symmetries.

\begin{figure}
    \centering
    \includegraphics[width=\columnwidth]{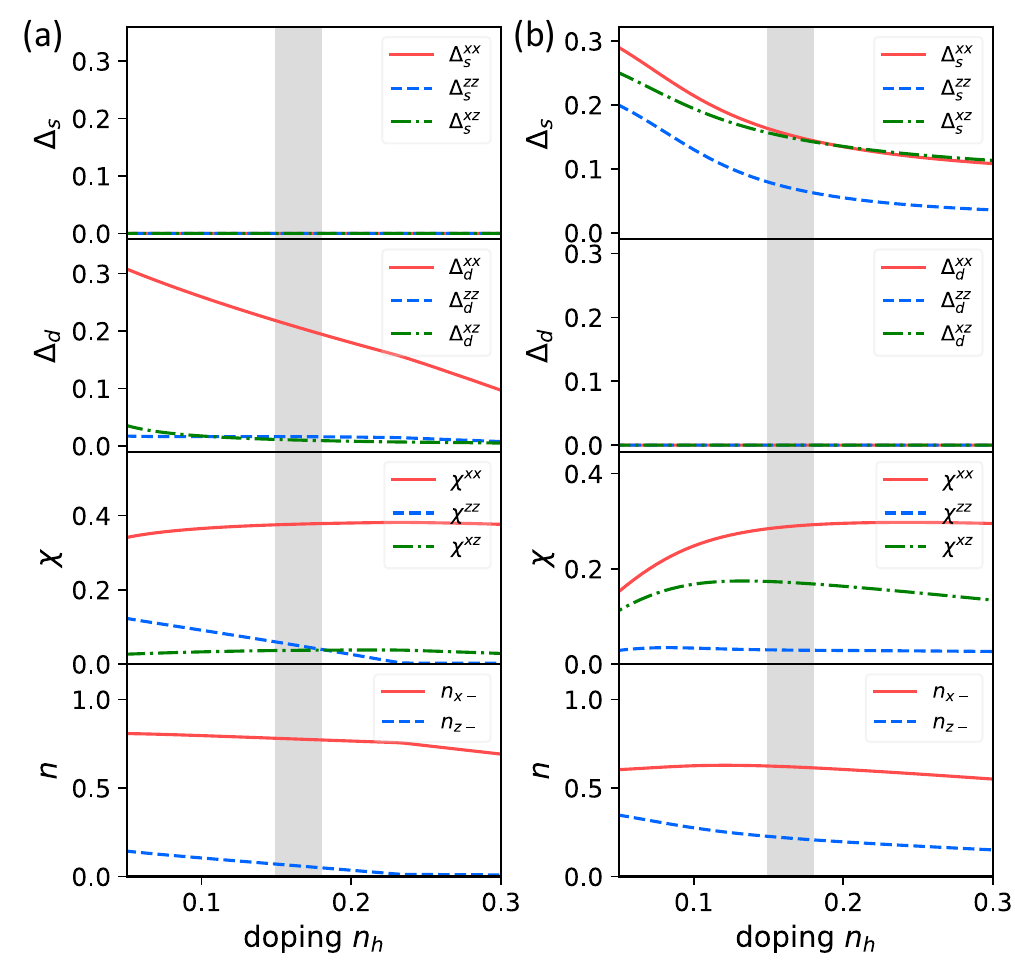}
    \caption{Mean field results for the two-orbital self-doped molecular Mott insulator obtained with in-plane intra-orbital hopping $t^{xx}=1$, $t^{zz}=0.2$, and inter-orbital hopping $t^{xz}=0.1$ (a), $t^{xz}=0.6$ (b), plotted as a function of the self-doped hole density $n_h$. The Hubbard interaction is set as $U_A=6$. The figures from top to bottom show the results for the pairing component in the extended s-wave $\Delta_{s}$ and d-wave $\Delta_{d}$, channel, the hopping mean field $\chi$ and the occupation number $n_{x-}$, $n_{z-}$. With increasing interorbital hybridization $t^{xz}$, we find the pairing symmetry changes from $d$-wave to $s$-wave. We estimate from DFT calculations~\cite{2025Jiang} $t^{xz}\approx0.4-0.6$ and self-doping $n_h\approx 0.17$ (area shaded in gray) corresponding to the $\alpha$-band.}
    \label{meanfield}
\end{figure}

The parameters in the calculations are chosen as follows. The Hubbard interaction $U_A=6$ and the nearest neighbor hoppings are set as $t^{xx}=1,\,t^{zz}=0.2$. As for interorbital hoppings, we consider two cases with $t^{xz}=0.1$ and $t^{xz}=0.6$. The calculations are performed as a function of self-doped hole density $n_h$. Each $n_h$ corresponds to a specific $\epsilon_{x+}$, e.g $n_h=0.1$ can be obtained with $\epsilon_{x+}\approx3$. The pairing symmetry can be categorized according to the representation of the lattice symmetry group $D_{4h}$. In the singlet channel, pairing between the nearest neighbors can belong to either the extended $s$-wave ($A_{1g}$ representation), denoted as $\Delta_s$, with the same sign in the $x$ and $y$ directions, or the $d$-wave ($B_{1g}$ representation), denoted as $\Delta_d$, with alternating signs. The mean field results are plotted in Fig.~\ref{meanfield}. For $t^{xz}=0.1$, as shown in Fig.~\ref{meanfield} (a), we find the pairing symmetry agrees with $d$-wave, similar to the single orbital $t-J$ model. In this case, the pairings are found to be dominated by $\Delta^{xx}$ (red solid line), i.e the intra-orbital pairing within the $|x,-\rangle$ orbital. $\Delta^{xz}$ (green dash-dot line) is found to be significantly smaller. $\Delta^{zz}$ is also small due to small particle number $n_z$, as shown in the lowest panel. For $t^{xz}=0.6$, as shown in Fig.~\ref{meanfield} (b), the pairing solution is found to agree with the extended $s$-wave. In this case, both intra- and interorbital pairings are of similar magnitude, suggesting strong interorbital effects. Moreover, the electron occupation number difference $n_{x-}-n_{z-}$ becomes much smaller as the interorbital hopping increases from $t^{xz}=0.1$ to $t^{xz}=0.6$, which also points to the enhanced  inter-orbital effects. We therefore conclude based on the above calculations that the extended $s$-wave pairing can be favored due to significant interorbital effect. For intermediate value of $t^{xz}$, the competition between $s$- and $d$-wave instability requires further study.

\section{Effect of $U$ and $J_H$ to molecule state}
\begin{figure}
    \centering
    \includegraphics[width=1\linewidth]{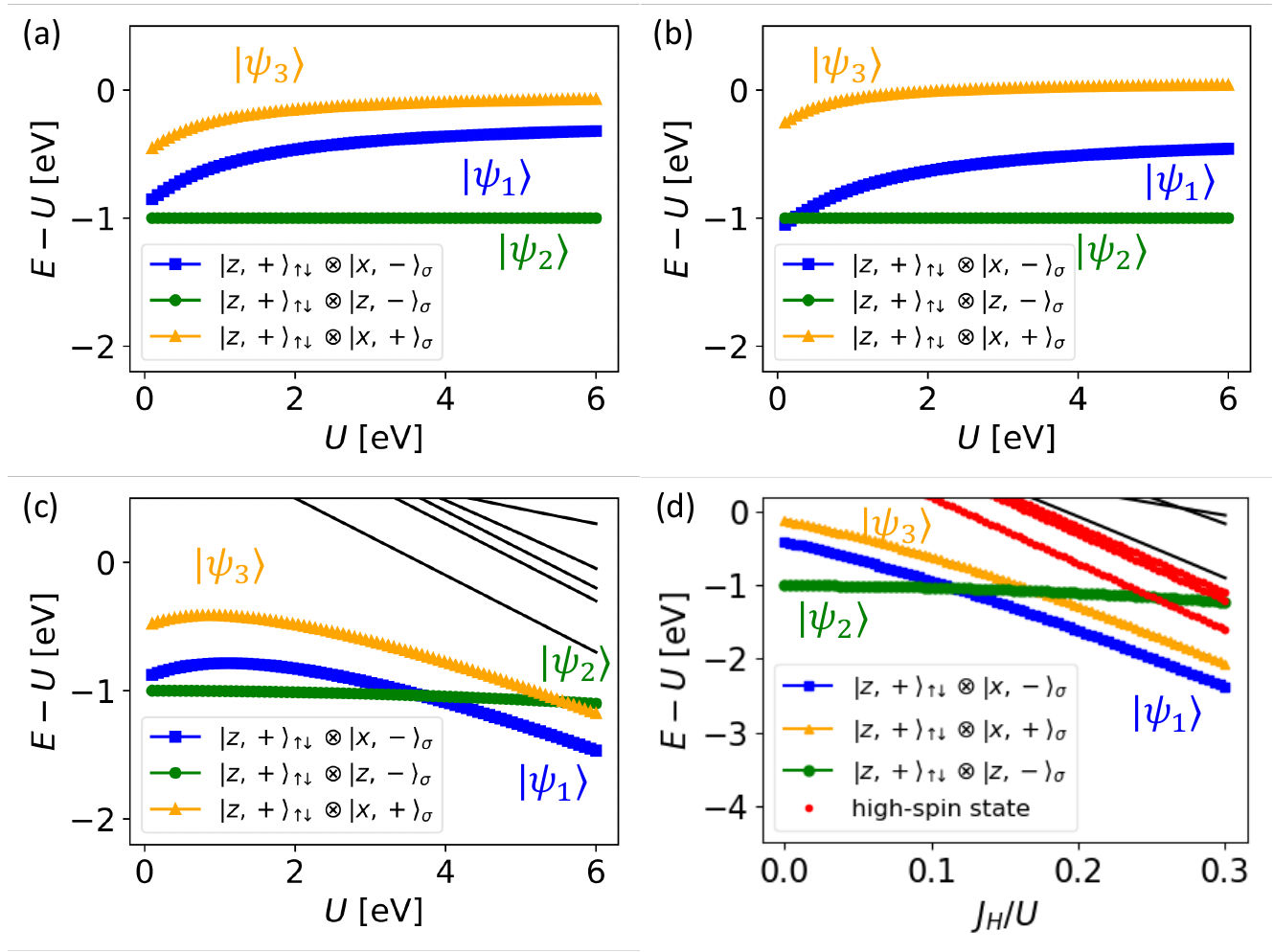}
    \caption{Local spectra and projected wave function components of the low-energy eigen states. (a) The spectra obtained under $t_\perp^z=-1\text{eV},\, t_\perp^x=0.2\text{eV},\, E_x=1.3\text{eV},\, J_H=0\text{eV}$. The dominant wave function components are plotted along the spectra of the three lowest energy eigenstates according to $|z,+\rangle_{\uparrow\downarrow} \otimes |x,-\rangle$ (blue), $|z,+\rangle_{\uparrow\downarrow} \otimes |z,-\rangle$ (green) and $|z,+\rangle_{\uparrow\downarrow} \otimes |x,+\rangle$ (yellow). The first two electronic configurations correspond to the two anti-symmetric orbitals forming the molecular Mott insulator, and the last one contributes to the self-doping effect. (b) The spectra obtained with $t_\perp^x=0.4\text{eV}$ in comparison to (a). The gap between the $|x,-\rangle $ and the $ |x,+\rangle $ becomes larger, indicating a weaker self-doping effect. (c) The spectra obtained with $J_H=0.1U$ in comparison to (a). The two eigenstates with electrons occupying different orbitals have smaller slope in $U$ due to Hund’s coupling. Around $U=4eV$, the local electronic spectra contains two nearly degenerate anti-symmetric orbitals $ |x,-\rangle$ and $ |z,-\rangle$, with $ |x,+\rangle $ above by $\sim t_\perp^x$. (d) The spectra obtained with $U=3$eV as a function of $J_H/U$. The molecular Mott insulating regime is valid below $J_H/U\sim0.25$. With larger $J_H/U$, the high spin states become energetically favored over the $|\psi_2\rangle$.}
    \label{model_params}
\end{figure}

In the previous section, we have shown that due to the crystal field splitting and the Jahn-Teller effect ,  the two sets of $e_g$ orbitals are relevant for the low energy physics of \LNO. The interlayer hoppings $t_\perp^{x,z}$ further introduce the molecular energy splitting between the interlayer (anti-)symmetric molecular orbitals. The effective low-energy theory in the large-$U$ limit is derived based on the local electronic structure. In this section, we will demonstrate that the low-energy orbitals we identified play the dominant role in the low-energy physics by performing exact diagonalization on a local molecular system composed of two Ni$^{2.5+}$ ions, one from the top layer and the other from the bottom layer. We analyze the impact of finite Hubbard repulsion $U$ and the role of interlayer hopping—especially the $t_\perp^x$ term, and we examine the influence of Hund's coupling, $J_H$. Consider a single Ni$_2^{5+}$ molecule with total $3$ electrons occupying 8 orbitals formed by the 2 sets of $e_g$ orbitals. The Hamiltonian for such a molecule can be written as:
\begin{align}
    H_{\text{molecule}}&=\sum_{\alpha\sigma} t_\perp^\alpha \Big( c_{t\alpha\sigma}^\dagger c_{b\alpha\sigma} + \text{h.c.}\Big)+ E_x\sum_{l\sigma} n_{x\sigma}\notag\\
    & +U\sum_{l\alpha}n_{l\alpha\uparrow}n_{l\alpha\downarrow}+U'\sum_l n_{lx}n_{lz}\\
    & +J_H\sum_l\sum_{\sigma\sigma'}c_{lx\sigma}^\dagger c_{lz\sigma'}^\dagger c_{lx\sigma'}c_{lz\sigma}
    +J_P\sum_l\sum_\alpha c_{l\alpha\uparrow}^\dagger c_{l\alpha\downarrow}^\dagger c_{l\bar{\alpha}\downarrow}c_{l\bar{\alpha}\uparrow}\notag
\end{align}
where $E_z=0$ is set for convenience. In the interacting part,  $J_H=J_P=J$ denotes Hund's coupling and the invariance under symmetry operations yields $U'=U-2J$. The hopping and interaction parameters here are associated with the atomic orbitals. The energy spectra are obtained through exact diagonalization of $H_{\text{molecule}}$ and the wave function contents of the eigen states are analyzed, as shown in Fig.~\ref{model_params}. The parameters here are $t_\perp^z=-1$eV and $E_x=1.3$eV as estimated from the DFT calculations in~\cite{2025Jiang} with a slightly larger $t^x_\perp=0.2$eV. 

For $J_H=0$, as depicted in Fig.~\ref{model_params} (a), the low-energy space contains three eigen states termed as $\{|\psi_1\rangle,\, |\psi_2\rangle,\, |\psi_3\rangle \}$. The eigenstates are usually found to be a hybridization of multiple electronic configurations. For the three lowest eigen states, we find $|\psi_1\rangle$ is dominated by $|z,+\rangle_{\uparrow\downarrow}\otimes|x,-\rangle_\sigma$ (blue), $|\psi_2\rangle$ is strictly $|z,+\rangle_{\uparrow\downarrow}\otimes|z,-\rangle_\sigma$ (green), and $|\psi_3\rangle$ is dominated by $|z,+\rangle_{\uparrow\downarrow}\otimes|x,+\rangle_\sigma$ (yellow). In particular, the overlap of $|\psi_{1,2,3}\rangle$ with the corresponding dominating states are found to be $\{\sim0.86,\,1,\,\sim0.85\}$ in the large-$U$ limit. As revealed by the dominating wave function components, the $|z,+\rangle$ orbital remains doubly-occupied, in agreement with the previous analysis that $|z,+\rangle$ lies well below other orbitals. Furthermore, At $U=0$, the two anti-symmetric orbitals, i.e $|x,-\rangle$ and $|z,-\rangle$, are nearly degenerate as reflected by $|\psi_{1,2}\rangle$ being close in the spectra. The $|x,+\rangle$ orbital lies above by $t_{\perp}^x$, leading to the self-doping effect. As $U$ increases, the energy of $|\psi_1\rangle$ and $|\psi_3\rangle$ orbital slightly increases due to the hybridization with other orbitals.

The effects of interlayer hopping $t_\perp^x$ is depicted in Fig.~\ref{model_params} (b) for $t_\perp^x=0.4$eV. As $t_\perp^x$ becomes larger,  the molecular gap between the $|x,+\rangle$ and $|x,-\rangle$ increases, as shown by the larger energy separation of $|\psi_1\rangle$ and $|\psi_3\rangle$ in Fig.~\ref{model_params} (b). The self-doping is therefore weakened with larger values of $t_\perp^x$.

The effects of $J_H$ are also of interest as studied in various works~\cite{wucj, 2025Si}. To demonstrate the effects of=Hund's coupling in the proposed molecular Mott insulator regime, we perform two sets of calculations as shown in Fig.~\ref{model_params} (c-d). Firstly, as depicted in Fig.~\ref{model_params} (c), the relative strength of Hund's coupling is fixed at $J_H=0.1U$ and the local electronic spectra is solved as a function of $U$. The low-energy space is similar to the previous two cases. However, the two states with electrons occupying different orbitals, i.e $|\psi_1\rangle$ and $|\psi_3\rangle$ are favored by Hund's coupling in the large-$U$ regime. In particular, for $U\sim4$eV according to the estimation of \LNO from first principle calculations~\cite{YXWang_PhysRevB.110.205122}, the electronic structure is in exact agreement with the proposed self-doped molecular Mott insulator. Next, we fix the value of $U=3$eV and study the energy spectra as a function of $J_H/U$. As shown in Fig.~\ref{model_params} (d), we find the molecular Mott insulator scenario is valid for $J_H/U\lesssim0.25$. As $J_H/U$ increases, the high spin states (red) become energetically favored over $|\psi_2\rangle$ (green).

\section{Summary}
In this paper, we start with two $e_g$ orbitals ($d_{x^2-y^2}$ and $d_{z^2}$) for double layer \LNO.  We consider interlayer coupling for $d_{z^2}$ orbitals to be strong and the two Ni-sites in the top and bottom layers form a molecule. The bonding states of $d_{z^2}$ are all occupied leaving one $e_g$ electron per molecule on average for the anti-bonding $d_{z^2}$ and both bonding and anti-bonding states of $d_{x^2-y^2}$. In the large limit of on-site Coulomb repulsion $U$, the system is described by the KK model or molecular Mott insulator with two orbitals (anti-bonding $d_{z^2}$ and bonding $d_{x^2-y^2}$) if all the anti-bonding states of $d_{x^2-y^2}$ are empty, and the system is described by self-doped molecular Mott insulator if the anti-bonding $d_{x^2-y^2}$ is slightly occupied. The latter may form a molecular doublon with the bonding $d_{x^2-y^2}$ orbital of the same spin, which does not cost $U$, hence providing holes in the background of the Mott insulator. We argue that the low energy physics is given by the doped two orbital $t-J$ model (more precisely KK model) on a square lattice.  Superconductivity arises from nearest neighbor orbital and spin coupling, whose pairing symmetry depends on the inter-orbital pairing strength. The effects of Hund's coupling is further analyzed based on the exact diagonalization calculations and the self-doped molecular Mott insulating regime is found to be valid under the realistic estimation of the model parameters. Our theory predicts more complex spin excitation in \LNO than in cuprates because of the two orbitals. Chemical hole doping, such as partial replacement of La by valence $2+$ ions (Ca or Sr) and interstitial oxygens, could introduce additional mobile holes hence good for superconductivity. For electron doping, the apical oxygen vacancies between bilayer Ni atoms destroy the molecular orbital, especially the anti-binding Fermi surface~\cite{2025Jiang}. Therefore, the oxygen vacancies play a similar role as impurities on CuO$_2$ planes in cuprates and severely damage the superconductivity.

\textit{Acknowledgement.}
The authors thank Hui-Ke Jin for helpful discussions. We acknowledge the support by the National Natural Science Foundation of China (Grant NSFC-12494590, No. NSFC-12174428), the Ministry of Science and Technology (Grant No. 2022YFA1403900), the Chinese Academy of Sciences Project for Young Scientists in Basic Research (2022YSBR-048), the Innovation program for Quantum Science and Technology (Grant No. 2021ZD0302500), and Chinese Academy of Sciences under contract No. JZHKYPT-2021-08.

\bibliography{reference}

\clearpage
\clearpage
\onecolumngrid
\begin{center}
\textbf{Supplemental Material for "Self-doped Molecular Mott Insulator for Bilayer High-Temperature Superconductivity"}
\end{center}

\setcounter{equation}{0}
\setcounter{figure}{0}
\setcounter{table}{0}
\setcounter{page}{1}
\makeatletter
\renewcommand{\theequation}{S\arabic{equation}}
\renewcommand{\thefigure}{S\arabic{figure}}
\renewcommand{\thetable}{S\arabic{table}}

\section{local electronic structure in the $U\rightarrow\infty$ limit} 
To analyze the local electronic structure, we consider a 2-site molecule formed by Ni$_2^{5+}$ in the $U\rightarrow\infty$ limit. The local Hamiltonian of the molecule reads:
\begin{equation}
    H_0=\sum_{\alpha\sigma} t_\perp^\alpha \Big( c_{t\alpha\sigma}^\dagger c_{b\alpha\sigma} + \text{h.c.}\Big)+\sum_{l=t,b} \sum_{\alpha}  U n_{l\alpha\uparrow} n_{l\alpha\downarrow}+\sum_{l=t,b}U^\prime n_{lx} n_{lz}+H_{onsite}.
\end{equation}
For consistency, we use $l=t,b$ to denote the two Ni$^{2.5+}$ ions from the two layers. The interlayer hybridization amplitude is denoted by $t_\perp^\alpha$, with $\alpha=x,z$ for the atomic orbital index. Note there is no inter-orbital hybridization due to the symmetry of the atomic orbital wave functions. We expect $|t_\perp^z|>|t_\perp^x|$ due to the significant wave function overlap between Ni $3d_{z^2}$ orbitals and the interlayer oxygen $p_z$ orbitals. As mentioned in the main text, one will find $t_\perp^z<0$ from the sign of the wave functions. For simplicity, we ignore the Hund's coupling and set $U = U^\prime$.

In addition, the Jahn-Teller effect is described by the orbital onsite energy
\begin{equation}
\label{honsite}
    H_{onsite} = E_x\sum_{l\sigma} n_{x\sigma} + E_z \sum_{l\sigma} n_{z\sigma}.
\end{equation}
Here $E_x>0$ and $E_z=0$ due to elongated apical oxygen bonds in the Ni-O octahedron.

For such a molecule, there are in total $3$ electrons occupying the $8$ local energy levels formed by the two sets of $e_g$ orbitals of the Ni$_2^{5+}$ ions. Therefore there are in total $56$ possible states. We begin by noting the symmetry and conserved quantities in $H_0$ in the local orbitals. We define the total spin polarization $s_z$ as the total spin along the $z$ direction and the orbital polarization $\tau_z=-1/2 (+1/2)$ for $z(x)$ orbital, respectively. The operators are defined as:
\begin{align}
    \hat{s}_z &= \frac12 \sum_{l,\alpha} (n_{l\alpha\uparrow} - n_{l\alpha\downarrow}); \\ 
    \hat{\tau}_z &= \frac 12 \sum_{l,\sigma} (n_{lx\sigma} - n_{lz\sigma}); \\
    \hat{s}_z \hat{\tau}_z &= \frac12 \sum_{l} (n_{lx\uparrow} - n_{lx\downarrow} - n_{lz\uparrow} + n_{lz\downarrow})
\end{align}
In the last line, we introduce a composite operator $\hat{s}_z \hat{\tau}_z$, describing the spin-orbital locking. All 3 operators commute with $H_0$ and give good quantum numbers denoted by $s_z,\tau_z,s_z\tau_z$, respectively.

We consider the local electronic states in the large $U$ limit and treat interlayer hopping in $H_\perp$ as small parameter. There are in total $56$ possible states, and $8$ of them contains $3$ electrons in the same layer, leading to total Hubbard repulsion of $3U$. The remaining $48$ states contains $2$ electrons in one layer and $1$ electron in the other layer, which corresponds to energy cost from Hubbard repulsion of $U$. In the low-energy limit, we consider the latter $48$ states and set $E_0=U$ as the energy reference point.

The $48$ states can be further categorized according to the three good quantum numbers $(s_z,\tau_z,s_z\tau_z)$ into $4$ groups and we calculate their energy eigen value resulting from interlayer hybridization $t_\perp^z$ ($t_\perp^x$ is set to $0$ here for simplicity).

Group I: $(s_z=\pm3/2,\tau_z=\pm1/2,s_z\tau_z=\pm1/2)$. In this case, the three electrons are of the same spin and occupy $3$ out of $4$ possible orbitals from the two layers, leading to $8$ states in total. The energy eigen values are $\pm t_\perp^z$ for $\tau_z=1/2$ (each is two-fold degenerate) and $0$ for $\tau_z=-1/2$ (four-fold degenerate).

Group II: $(s_z=\pm1/2,\tau_z=\pm3/2,s_z\tau_z=\pm1/2)$. In this case, the three electrons are orbital-polarized. There are $8$ states in total. The energy eigen values are $\pm t_\perp^z$ for $\tau_z=-3/2$ (each is two-fold degenerate) and $0$ for $\tau_z=3/2$ (four-fold degenerate).

Group III: $(s_z=\pm1/2,\tau_z=\pm1/2,s_z\tau_z=\pm3/2)$. In this case, the spin and orbital are locked, e.g the case with all spin-up electrons in $x$ orbital and all spin-down electrons in $z$ orbital corresponds to $s_z\tau_z=3/2$. There are in total $8$ states belonging to this sector. The energy eigen values are $\pm t_\perp^z$ for $\tau_z=1/2$ (each is two-fold degenerate) and $0$ for $\tau_z=-1/2$ (four-fold degenerate).

Group IV: $(s_z=\pm1/2,\tau_z=\pm1/2,s_z\tau_z=\pm1/2)$. Here none of the three conserved quantities are fully-polarized. There are $24$ states belonging to this group. The $24$ states can be further divided into $4$ subgroups which are labeled as $ (s_z,\tau_z,s_z\tau_z)\in \Big\{(\frac12,-\frac12,\frac12), (-\frac12,-\frac12,-\frac12), (\frac12,\frac12,-\frac12), (-\frac12,\frac12,\frac12)\Big\}$. Take the first subgroup as an example, the $6$ states can be denoted as:
\begin{align}
    |\psi_1\rangle = c_{bx\uparrow}^\dagger c_{tz\uparrow}^\dagger c_{tz\downarrow}^\dagger |\text{vac}\rangle, \quad &|\psi_2\rangle = c_{tx\uparrow}^\dagger c_{bz\uparrow}^\dagger c_{bz\downarrow}^\dagger |\text{vac}\rangle, \notag\\
    |\psi_3\rangle = c_{bx\uparrow}^\dagger c_{tz\uparrow}^\dagger c_{bz\downarrow}^\dagger |\text{vac}\rangle, \quad &|\psi_4\rangle = c_{tx\uparrow}^\dagger c_{tz\uparrow}^\dagger c_{bz\downarrow}^\dagger |\text{vac}\rangle, \notag\\
    |\psi_5\rangle = c_{bx\uparrow}^\dagger c_{tz\downarrow}^\dagger c_{bz\uparrow}^\dagger |\text{vac}\rangle, \quad &|\psi_6\rangle = c_{tx\uparrow}^\dagger c_{tz\downarrow}^\dagger c_{bz\uparrow}^\dagger |\text{vac}\rangle \notag.
\end{align}
In writing the interlayer hybridization, we choose the following basis for convenience:
\begin{align}
    |1\rangle &= \frac{1}{\sqrt2} (c_{bx\uparrow}^\dagger c_{tz\uparrow}^\dagger c_{tz\downarrow}^\dagger + c_{tx\uparrow}^\dagger c_{bz\uparrow}^\dagger c_{bz\downarrow}^\dagger) |\text{vac}\rangle, \notag\\ 
    |2\rangle &= \frac{1}{\sqrt2} (c_{bx\uparrow}^\dagger c_{tz\uparrow}^\dagger c_{tz\downarrow}^\dagger - c_{tx\uparrow}^\dagger c_{bz\uparrow}^\dagger c_{bz\downarrow}^\dagger) |\text{vac}\rangle, \notag\\
    |3\rangle &= \frac{1}{2} (c_{bx\uparrow}^\dagger + c_{tx\uparrow}^\dagger) (c_{tz\uparrow}^\dagger c_{bz\downarrow}^\dagger - c_{tz\downarrow}^\dagger c_{bz\uparrow}^\dagger) |\text{vac}\rangle, \notag\\
    |4\rangle &= \frac{1}{2} (c_{bx\uparrow}^\dagger - c_{tx\uparrow}^\dagger) (c_{tz\uparrow}^\dagger c_{bz\downarrow}^\dagger - c_{tz\downarrow}^\dagger c_{bz\uparrow}^\dagger) |\text{vac}\rangle \notag\\
    |5\rangle &= \frac{1}{2} (c_{bx\uparrow}^\dagger + c_{tx\uparrow}^\dagger) (c_{tz\uparrow}^\dagger c_{bz\downarrow}^\dagger + c_{tz\downarrow}^\dagger c_{bz\uparrow}^\dagger) |\text{vac}\rangle, \notag\\
    |6\rangle &= \frac{1}{2} (c_{bx\uparrow}^\dagger - c_{tx\uparrow}^\dagger) (c_{tz\uparrow}^\dagger c_{bz\downarrow}^\dagger + c_{tz\downarrow}^\dagger c_{bz\uparrow}^\dagger) |\text{vac}\rangle \notag\\
\end{align}
Note the states $|5\rangle$ and $|6\rangle$ don't hybridize with other states in this subgroup under interlayer hopping $t^\alpha_\perp$ and their energy eigen values are given by $E_5=t_\perp^x,\,E_6=-t_\perp^x$. Under the basis of the first $4$ states, the Hamiltonian associated with $t^\alpha_\perp$ can be written as:
\begin{equation}
    H_\perp=\left(\begin{array}{cccc}
         0 & 0 & \sqrt2t_\perp^z & 0 \\
         0 & 0 & 0 & \sqrt2t_\perp^z  \\
         \sqrt2t_\perp^z &  0 & t_\perp^x & 0 \\
         0 & \sqrt2t_\perp^z & 0 & -t_\perp^x
    \end{array}\right).
\end{equation}
For $t_\perp^x=0$, this subgroup has the lowest energy eigen value $-\sqrt2t_\perp^z$, which is lower than the minimum of other groups $-t_\perp^z$. With a nonzero $t_\perp^x$, the energy eigen value can be obtained as:
\begin{align}
    E_1^\pm &= 
    \pm\frac12\Big(t_\perp^x+\sqrt{(t_\perp^x)^2+8(t_\perp^z)^2}\Big),\\
    E_2^\pm &= 
    \pm\frac12\Big(t_\perp^x-\sqrt{(t_\perp^x)^2+8(t_\perp^z)^2}\Big).
\end{align}

Consider the case with $E_x-E_z\approx |t_\perp^z|$ and $t_\perp^z<0$ as mentioned in the main text, the low-energy space is spanned by the following three states:
\begin{align}
    |\psi_1\rangle&= a_1|z,+\rangle_{\uparrow\downarrow}\otimes|x,-\rangle_\sigma+\text{other components},\quad \lambda_1=-\frac12\Big(t_\perp^x+\sqrt{(t_\perp^x)^2+8(t_\perp^z)^2}\Big)+E_x,\notag\\
    |\psi_2\rangle&= |z,+\rangle_{\uparrow\downarrow}\otimes|z,-\rangle_\sigma,\quad \lambda_2=t_\perp^z,\notag\\
    |\psi_3\rangle&= a_3|z,+\rangle_{\uparrow\downarrow}\otimes|x,+\rangle_\sigma+\text{other components},\quad \lambda_3=\frac12\Big(t_\perp^x-\sqrt{(t_\perp^x)^2+8(t_\perp^z)^2}\Big)+E_x.\notag
\end{align}
Here $\lambda$ denotes the energy eigen value of $H_0$ with respect to the reference point $E_0=U$. For $E_x=0.8$, $t_\perp^z=-1$ and $t_\perp^x=0.2$, the three eigen energies are given by $\{\lambda_1=-0.72,\,\lambda_2=-1,\,\lambda_3=-0.52\}$. The state $|z,+\rangle_{\uparrow\downarrow}\otimes|z,-\rangle_\sigma$ is exactly the eigenstate $|\psi_2\rangle$. The other two eigenstates $|\psi_1\rangle$ and $|\psi_3\rangle$ are dominated by their first components with $a_1\approx0.86$ and $a_3\approx0.85$.

\section{renormalized mean field calculation}
The effective model in given by eq.~\eqref{model} in the main text:
\begin{align}
    H_{\text{eff}}&=H_S+H_{A,t}+H_{A,int}
\end{align}

The hopping part of the bonding electrons from $|x,+\rangle$ orbital is described by eq.~\eqref{H1}:
\begin{equation}
    H_S=-\sum_{\langle ij\rangle,\sigma}t_{ij}^+(c_{i,x+,\sigma}^\dagger c_{j,x+,\sigma}+\text{h.c.})+\epsilon_{x+}\sum_{i\sigma}c_{i,x+,\sigma}^\dagger c_{i,x+,\sigma}, 
\end{equation}
where we use $c_{i,x+,\sigma}^\dagger$ to denote the creation operator corresponding to $|x,+\rangle_\sigma$ with spin $\sigma$ and $\epsilon_{x+}>0$ denotes the molecular energy shift of the bonding orbital.

The hopping part of the antisymmetric electrons from $|x,-\rangle$ and $|z,-\rangle$ orbitals is given by eq.~\eqref{H0}:
\begin{align}
    H_{A,t}&=-\sum_{\langle ij\rangle,\alpha\alpha^\prime}t_{ij}^{\alpha\alpha^\prime} \hat{P}_G(c_{i,\alpha-,\sigma}^\dagger c_{j,\alpha^\prime-,\sigma}+\text{h.c.})\hat{P}_G\notag\\
    &=-\sum_{\langle ij\rangle,\alpha\alpha^\prime}g_{t}^{\alpha\alpha^\prime}t_{ij}^{\alpha\alpha^\prime} (c_{i,\alpha-,\sigma}^\dagger c_{j,\alpha^\prime-,\sigma}+\text{h.c.})
\end{align}
Here $c_{i,\alpha-,\sigma}^\dagger$ with $\alpha,\alpha^\prime=x,z$ is the creation operator associated with $|x,-\rangle$ and $|z,-\rangle$. $\hat{P}_G$ in the first line denotes the projection operation from the double-occupancy constraint in the large-$U$ limit. The Gutzwiller factor is defined as: $g_t^{\alpha\alpha^\prime}=\frac{2n_h}{\sqrt{(2-n_{\alpha-})(2-n_{\alpha^\prime-})}}$, where $n_h=n_{x+}$ denotes the number of self-doped holes, and $n_{\alpha-}$ is the total electron number from the orbital $|x,-\rangle$ and $|z,-\rangle$.

The effective interaction term of the antisymmetric electrons is derived in eq.~\eqref{Hint}:
\begin{align}
    H_{A,int}=&\frac{4}{U_A}\sum_{\langle ij\rangle}\sum_{\alpha\alpha^\prime\beta\beta^\prime}
\left[t_{ij}^{\alpha\beta}t_{ij}^{\alpha^\prime\beta^\prime}\bm{S}_{i\alpha\alpha^\prime} \cdot \bm{S}_{j\beta^\prime\beta} - \frac14 \Big(t_{ij}^{\alpha\beta}t_{ij}^{\alpha^\prime\beta^\prime} -   t_{ij}^{\alpha\bar{\beta^\prime}}t_{ij}^{\alpha^\prime\bar{\beta}}(-1)^{\delta_{\beta\beta^\prime}} - t_{ij}^{\bar{\alpha}\beta^\prime}t_{ij}^{\bar{\alpha^\prime}\beta }(-1)^{\delta_{\alpha\alpha^\prime}} \Big)n_{i\alpha\alpha^\prime} n_{j \beta^\prime\beta}\right].
\end{align}
The spin operator is defined as $\bm{S}_{i\alpha\beta}=\frac12 c_{i,\alpha-}^\dagger \bm{\sigma} c_{i,\beta-}$ and the density operator $n_{i\alpha\beta}=c_{i,\alpha-}^\dagger c_{i,\beta-}$, with $c_{i,\beta-}=(c_{i,\beta-,\uparrow}, \, c_{i,\beta-,\downarrow})^T$. To decouple the two quartic terms in the interaction Hamiltonian, we introduce two types of mean fields. One is the singlet pairing mean field $\Delta_{ij}^{\alpha\alpha^\prime}$ as in eq.~\eqref{pairing}:
\begin{equation}
    \Delta_{ij}^{\alpha\alpha^\prime}=\langle c_{i,\alpha-,\uparrow}^\dagger c_{j,\alpha^\prime-,\downarrow}^\dagger\rangle-\langle c_{i,\alpha-,\downarrow}^\dagger c_{j,\alpha^\prime-,\uparrow}^\dagger\rangle.
\end{equation}
The other one is the hopping mean field $\chi_{ij}^{\alpha\alpha^\prime}$:
\begin{eqnarray}
    \chi_{ij}^{\alpha\alpha^\prime}=\langle c_{i,\alpha-,\uparrow}^\dagger c_{j,\alpha^\prime-,\uparrow}\rangle+\langle c_{i,\alpha-,\downarrow}^\dagger c_{j,\alpha^\prime-,\downarrow}\rangle.
\end{eqnarray}

This leads to the following form of the mean field Hamiltonian, written in matrix form under the basis $\Psi_{\bm{k}}=(c_{\bm{k},x+,\uparrow},c_{\bm{k},x-,\uparrow},c_{\bm{k},z-,\uparrow},c_{-\bm{k},x+,\downarrow}^\dagger,c_{-\bm{k},x-,\downarrow}^\dagger,c_{-\bm{k},z-,\downarrow}^\dagger)^T$:
\begin{eqnarray}
    H_{MF}=\sum_{\bm{k}}\Psi_{\bm{k}}^\dagger H_{MF}^{\bm{k}}\Psi_{\bm{k}}=\sum_{\bm{k}}\Psi_{\bm{k}}^\dagger\left(\begin{array}{cccccc}
         \epsilon_{\bm{k}}^{x+}-\mu & & & & & \\
         & \epsilon_{\bm{k}}^{xx}-\mu & \epsilon_{\bm{k}}^{xz} & & \Delta_{\bm{k}}^{xx,*} & \Delta_{\bm{k}}^{xz,*} \\
         & \epsilon_{\bm{k}}^{zx} & \epsilon_{\bm{k}}^{zz}-\mu & & \Delta_{\bm{k}}^{zx,*} & \Delta_{\bm{k}}^{zz,*} \\
         & & & -\epsilon_{-\bm{k}}^{x+}+\mu & & \\
         & \Delta_{\bm{k}}^{xx} & \Delta_{\bm{k}}^{xz} & & -\epsilon_{-\bm{k}}^{xx}+\mu & -\epsilon_{-\bm{k}}^{xz} \\
         & \Delta_{\bm{k}}^{zx} & \Delta_{\bm{k}}^{zz} & & -\epsilon_{-\bm{k}}^{zx} & -\epsilon_{-\bm{k}}^{zz}+\mu
    \end{array}\right)\Psi_{\bm{k}}.
\end{eqnarray}
Here the chemical potential is denoted by $\mu$. The matrix elements are written as follows. For kinetic energy of the bonding orbital:
\begin{equation}
    \epsilon_{\bm{k}}^{x+}=-\sum_{ij}t_{ij}^{+}e^{-i\bm{k}\cdot\bm{r}_{ij}}.
\end{equation}
Here we use the sum over $ij$ without $\langle...\rangle$ to denote the site sum over site indices $ij$. The vector connecting $ij$ is denoted as $\bm{r}_{ij}=\bm{r}_i-\bm{r}_j$. For the two antisymmetric orbitals:
\begin{align}
    \epsilon_{\bm{k}}^{\alpha\alpha^\prime}=-\sum_{ ij}g_t^{\alpha\alpha^\prime}t_{ij}^{\alpha\alpha^\prime}e^{-i\bm{k}\cdot\bm{r}_{ij}}
    &-\frac{1}{2U_A}\sum_{\beta}\sum_{ ij}\left(t_{ij}^{\beta\alpha^\prime}t_{ij}^{\alpha\alpha^\prime}\chi_{ij}^{\beta\alpha^\prime}-
    t_{ij}^{\beta\bar{\alpha^\prime}}t_{ij}^{\alpha\bar{\alpha^\prime}}\chi_{ij}^{\beta\alpha^\prime}+2t_{ij}^{\beta\bar{\alpha^\prime}}t_{ij}^{\alpha\alpha^\prime}\chi_{ij}^{\beta\bar{\alpha^\prime}}\right)e^{-i\bm{k}\cdot\bm{r}_{ij}}\notag\\
    &-\frac{1}{2U_A}\sum_{\beta}\sum_{ ij}\left(t_{ij}^{\beta\alpha}t_{ij}^{\alpha^\prime\alpha}\chi_{ij}^{\beta\alpha}
    -t_{ij}^{\beta\bar{\alpha}}t_{ij}^{\alpha^\prime\bar{\alpha}}\chi_{ij}^{\beta\alpha}
    +2t_{ij}^{\beta\bar{\alpha}}t_{ij}^{\alpha^\prime\alpha}\chi_{ij}^{\beta\bar{\alpha}}\right)e^{i\bm{k}\cdot\bm{r}_{ij}}.
\end{align}
For pairings:
\begin{align}
    \Delta_{\bm{k}}^{\alpha\alpha^\prime}=
    &-\frac{1}{2U_A}\sum_{\beta}\sum_{ ij}\left(2t_{ij}^{\beta\alpha^\prime}t_{ij}^{\alpha\alpha^\prime}\Delta_{ij}^{\beta\alpha^\prime}
    +t_{ij}^{\beta\alpha^\prime}t_{ij}^{\alpha\bar{\alpha^\prime}}\Delta_{ij}^{\beta\bar{\alpha^\prime}}
    +t_{ij}^{\beta\bar{\alpha^\prime}}t_{ij}^{\alpha\bar{\alpha^\prime}}\Delta_{ij}^{\beta\alpha^\prime}\right)e^{-i\bm{k}\cdot\bm{r}_{ij}}\notag\\
    &-\frac{1}{2U_A}\sum_{\beta}\sum_{ ij}\left(2t_{ij}^{\beta\alpha}t_{ij}^{\alpha^\prime\alpha}\Delta_{ij}^{\beta\alpha}
    +t_{ij}^{\beta\alpha}t_{ij}^{\alpha^\prime\bar{\alpha}}\Delta_{ij}^{\beta\bar{\alpha}}
    +t_{ij}^{\beta\bar{\alpha}}t_{ij}^{\alpha^\prime\bar{\alpha}}\Delta_{ij}^{\beta\alpha}\right)e^{i\bm{k}\cdot\bm{r}_{ij}}.
\end{align}
Here the orbital indices $\alpha\alpha^\prime\beta$ can take either $x$ or $z$. The $\bar{\alpha}$ with the bar refers to the other orbital.

The Hamiltonian can be diagonalized by a unitary transformation $U^{\bm{k}}$ to get the Bogoliubov-de-Gennes quasiparticle dispersion $E_{n\bm{k}}$ and the corresponding quasi-particles $d_{n\bm{k}}$:
\begin{equation}
    \Lambda_{\bm{k}}=U^{\bm{k}\dagger}H_{MF}^{\bm{k}} U^{\bm{k}},
\end{equation}
where $\Lambda_{\bm{k}}$ is the diagonal matrix of the quasiparticle energy eigenvalue $E_{n\bm{k}}$. The self-consistent equations can be written in terms of the matrix elements of $U^{\bm{k}}$, denoted as $u_{in}^{\bm{k}}$. For pairing mean fields:
\begin{align}
    \Delta_{ij}^{xx}&=\langle c_{i,x-,\uparrow}^\dagger c_{j,x-,\downarrow}^\dagger\rangle-\langle c_{i,x-,\downarrow}^\dagger c_{j,x-,\uparrow}^\dagger\rangle=\frac{1}{N}\sum_{\bm{k}}\left[e^{-i\bm{k}\cdot\bm{r}_{ij}}\sum_nu_{2n}^{\bm{k}*}u_{5n}^{\bm{k}}n_F(E_{n\bm{k}})
    +e^{i\bm{k}\cdot\bm{r}_{ij}}\sum_nu_{2n}^{\bm{k}*}u_{5n}^{\bm{k}}n_F(E_{n\bm{k}})\right],\notag\\
    \Delta_{ij}^{xz}&=\langle c_{i,x-,\uparrow}^\dagger c_{j,z-,\downarrow}^\dagger\rangle-\langle c_{i,x-,\downarrow}^\dagger c_{j,z-,\uparrow}^\dagger\rangle=\frac{1}{N}\sum_{\bm{k}}\left[e^{-i\bm{k}\cdot\bm{r}_{ij}}\sum_nu_{2n}^{\bm{k}*}u_{6n}^{\bm{k}}n_F(E_{n\bm{k}})
    +e^{i\bm{k}\cdot\bm{r}_{ij}}\sum_nu_{3n}^{\bm{k}*}u_{5n}^{\bm{k}}n_F(E_{n\bm{k}})\right]=\Delta_{ij}^{zx},\notag\\
    \Delta_{ij}^{zz}&=\langle c_{i,z-,\uparrow}^\dagger c_{j,z-,\downarrow}^\dagger\rangle-\langle c_{i,z-,\downarrow}^\dagger c_{j,z-,\uparrow}^\dagger\rangle=\frac{1}{N}\sum_{\bm{k}}\left[e^{-i\bm{k}\cdot\bm{r}_{ij}}\sum_nu_{3n}^{\bm{k}*}u_{6n}^{\bm{k}}n_F(E_{n\bm{k}})
    +e^{i\bm{k}\cdot\bm{r}_{ij}}\sum_nu_{3n}^{\bm{k}*}u_{6n}^{\bm{k}}n_F(E_{n\bm{k}})\right].\notag
\end{align}
Here $n_F(E_{n\bm{k}})$ denotes the Fermi-Dirac distribution of the quasiparticle with energy $E_{n\bm{k}}$. The last equality of the second line comes from the ansatz that the pairing is an even function of the orbital index. For hopping mean fields:
\begin{align}
    \chi_{ij}^{xx}&=\langle c_{i,x-,\uparrow}^\dagger c_{j,x-,\uparrow}\rangle+\langle c_{i,x-,\downarrow}^\dagger c_{j,x-,\downarrow}\rangle=\frac{1}{N}\sum_{\bm{k}}\left[e^{-i\bm{k}\cdot\bm{r}_{ij}}\sum_nu_{2n}^{\bm{k}*}u_{2n}^{\bm{k}}n_F(E_{n\bm{k}})
    +e^{i\bm{k}\cdot\bm{r}_{ij}}\sum_nu_{5n}^{\bm{k}}u_{5n}^{\bm{k}*}(1-n_F(E_{n\bm{k}}))\right],\notag\\
    \chi_{ij}^{xz}&=\langle c_{i,x-,\uparrow}^\dagger c_{j,z-,\uparrow}\rangle+\langle c_{i,x-,\downarrow}^\dagger c_{j,z-,\downarrow}\rangle=\frac{1}{N}\sum_{\bm{k}}\left[e^{-i\bm{k}\cdot\bm{r}_{ij}}\sum_nu_{2n}^{\bm{k}*}u_{3n}^{\bm{k}}n_F(E_{n\bm{k}})
    +e^{i\bm{k}\cdot\bm{r}_{ij}}\sum_nu_{5n}^{\bm{k}}u_{6n}^{\bm{k}*}(1-n_F(E_{n\bm{k}}))\right],\notag\\
    \chi_{ij}^{zz}&=\langle c_{i,z-,\uparrow}^\dagger c_{j,z-,\uparrow}\rangle+\langle c_{i,z-,\downarrow}^\dagger c_{j,z-,\downarrow}\rangle=\frac{1}{N}\sum_{\bm{k}}\left[e^{-i\bm{k}\cdot\bm{r}_{ij}}\sum_nu_{3n}^{\bm{k}*}u_{3n}^{\bm{k}}n_F(E_{n\bm{k}})
    +e^{i\bm{k}\cdot\bm{r}_{ij}}\sum_nu_{6n}^{\bm{k}}u_{6n}^{\bm{k}*}(1-n_F(E_{n\bm{k}}))\right].\notag
\end{align}
And the particle number can be calculated according to:
\begin{align}
    n_{x+}&=\langle c_{i,x+,\uparrow}^\dagger c_{i,x+,\uparrow}\rangle+\langle c_{i,x+,\downarrow}^\dagger c_{i,x+,\downarrow}\rangle=\frac{1}{N}\sum_{n\bm{k}}\left[u_{1n}^{\bm{k}*}u_{1n}^{\bm{k}}n_F(E_{n\bm{k}})
    +u_{4n}^{\bm{k}}u_{4n}^{\bm{k}*}(1-n_F(E_{n\bm{k}}))\right],\notag\\
    n_{x-}&=\langle c_{i,x-,\uparrow}^\dagger c_{i,x-,\uparrow}\rangle+\langle c_{i,x-,\downarrow}^\dagger c_{i,x-,\downarrow}\rangle=\frac{1}{N}\sum_{n\bm{k}}\left[u_{2n}^{\bm{k}*}u_{2n}^{\bm{k}}n_F(E_{n\bm{k}})
    +u_{5n}^{\bm{k}}u_{5n}^{\bm{k}*}(1-n_F(E_{n\bm{k}}))\right],\notag\\
    n_{z-}&=\langle c_{i,z-,\uparrow}^\dagger c_{i,z-,\uparrow}\rangle+\langle c_{i,z-,\downarrow}^\dagger c_{i,z-,\downarrow}\rangle=\frac{1}{N}\sum_{n\bm{k}}\left[u_{3n}^{\bm{k}*}u_{3n}^{\bm{k}}n_F(E_{n\bm{k}})
    +u_{6n}^{\bm{k}}u_{6n}^{\bm{k}*}(1-n_F(E_{n\bm{k}}))\right].\notag
\end{align}

In calculations, the parameters are chosen as follows. The Hubbard interaction $U_A=6$ and the in-plane nearest neighbor hoppings are given by $t^{xx}=1,\,t^{zz}=0.2$ according to~\cite{dagotto2} for the case of \LNO. The interorbital hoppings $t^{xz}=0.1$ and $t^{xz}=0.8$ are considered. Moreover, the calculations are performed with varying number of self-doped holes $n_h$, which corresponds to varying the molecular energy splitting $\epsilon_{x+}$. The total electron number is fixed by the chemical potential $\mu$ to yield $\langle n\rangle=\langle n_h\rangle+\langle n_{x-}\rangle+\langle n_{z-}\rangle=1$ without external holes.

\subsection{pairing interaction from the two-orbital process}
\begin{figure}
    \centering
    \includegraphics[width=0.5\linewidth]{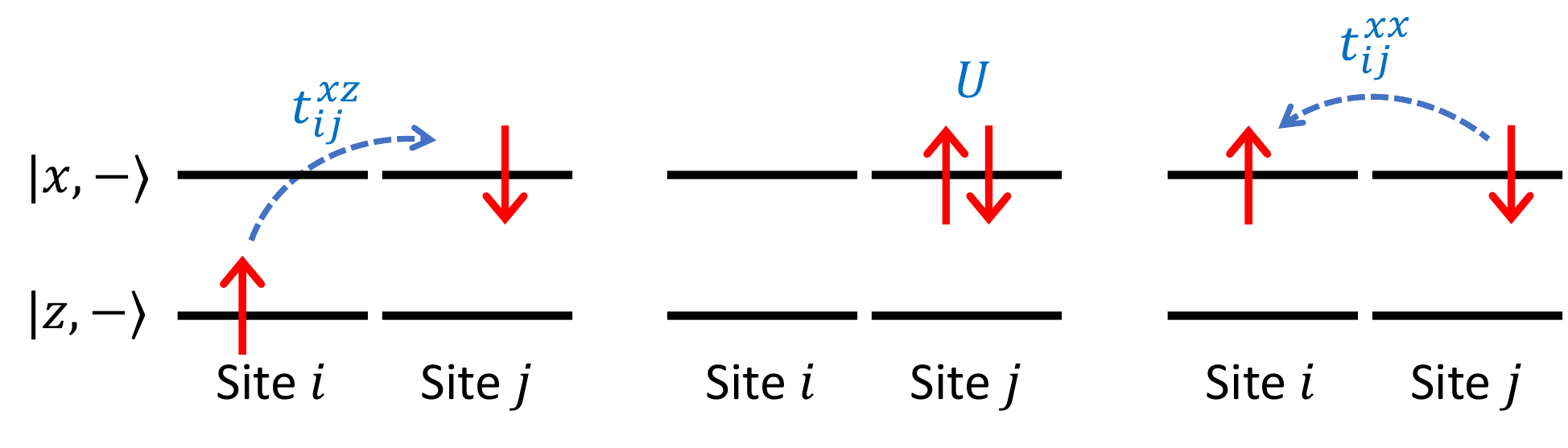}
    \caption{Virtual hopping process corresponding to the interaction term $c_{ix\uparrow}^\dagger c_{jx\downarrow}^\dagger c_{jx\downarrow} c_{iz\uparrow}$, leading to the hybrid pairing interaction of the form $\Delta^{xx,\dagger} \Delta^{xz}$ with interaction strength $\propto t_{ij}^{xz}t_{ij}^{xx}/{U_A}$.}
    \label{virtual_hybrid}
\end{figure}

The pairing part of the Hamiltonian takes the form:
\begin{eqnarray}
    H_{\text{pairing}}=\sum_{ij}\sum_{\alpha\alpha^\prime\beta}\left[-\frac{t_{ij}^{\alpha\beta}t_{ij}^{\alpha^\prime\beta}}{2U_A}\left(2\Delta_{ij}^{\alpha^\prime\beta,\dagger}\Delta_{ij}^{\alpha\beta}+\Delta_{ij}^{\alpha^\prime\bar{\beta},\dagger}\Delta_{ij}^{\alpha\bar{\beta}}\right)-\frac{t_{ij}^{\alpha\beta}t_{ij}^{\alpha^\prime\bar{\beta}}}{2U_A}\Delta_{ij}^{\alpha^\prime\beta,\dagger}\Delta_{ij}^{\alpha\bar{\beta}}\right].
\end{eqnarray}
By further considering the orbital indices, the pairing part can be divided into the intra-orbital, inter-orbital as well as hybrid channels (sum over $\alpha$ is implied):
\begin{align}
    H_{\text{pairing}}^{\text{intra-orbital}}&=-\frac{t_{ij}^{\alpha\alpha}t_{ij}^{\alpha\alpha}}{U_A}\Delta_{ij}^{\alpha\alpha,\dagger}\Delta_{ij}^{\alpha\alpha}
    -\frac{t_{ij}^{\alpha\bar{\alpha}}t_{ij}^{\alpha\bar{\alpha}}}{2U_A}\Delta_{ij}^{\alpha\alpha,\dagger}\Delta_{ij}^{\alpha\alpha}
    -\frac{t_{ij}^{\bar{\alpha}\alpha}t_{ij}^{\alpha\bar{\alpha}}}{2U_A}\Delta_{ij}^{\alpha\alpha,\dagger}\Delta_{ij}^{\bar{\alpha}\bar{\alpha}},\\
    H_{\text{pairing}}^{\text{inter-orbital}}&=-\frac{t_{ij}^{\alpha\bar{\alpha}}t_{ij}^{\alpha\bar{\alpha}}}{U_A}\Delta_{ij}^{\alpha\bar{\alpha},\dagger}\Delta_{ij}^{\alpha\bar{\alpha}}
    -\frac{t_{ij}^{\alpha\alpha}t_{ij}^{\alpha\alpha}}{2U_A}\Delta_{ij}^{\alpha\bar{\alpha},\dagger}\Delta_{ij}^{\alpha\bar{\alpha}}
    -\frac{t_{ij}^{\alpha\alpha}t_{ij}^{\bar{\alpha}\bar{\alpha}}}{2U_A}\Delta_{ij}^{\bar{\alpha}\alpha,\dagger}\Delta_{ij}^{\alpha\bar{\alpha}},\\
    H_{\text{pairing}}^{\text{hybrid}}&=-\frac{t_{ij}^{\alpha\alpha}t_{ij}^{\bar{\alpha}\alpha}}{U_A}\Delta_{ij}^{\bar{\alpha}\alpha,\dagger}\Delta_{ij}^{\alpha\alpha}
    -\frac{t_{ij}^{\alpha\alpha}t_{ij}^{\bar{\alpha}\alpha}}{2U_A}\Delta_{ij}^{\bar{\alpha}\bar{\alpha},\dagger}\Delta_{ij}^{\alpha\bar{\alpha}}
    -\frac{t_{ij}^{\alpha\bar{\alpha}}t_{ij}^{\alpha\alpha}}{2U_A}\Delta_{ij}^{\alpha\bar{\alpha},\dagger}\Delta_{ij}^{\alpha\alpha}+\text{h.c.}.
\end{align}
The last equation is unique to the considered two orbital interactions. Taking the first term in the last line, $-\frac{t_{ij}^{\alpha\alpha}t_{ij}^{\bar{\alpha}\alpha}}{U_A}\Delta_{ij}^{\bar{\alpha}\alpha,\dagger}\Delta_{ij}^{\alpha\alpha}$, and set $\alpha=x,\,\bar{\alpha}=z$ as an example, the virtual process giving this term can be depicted in Fig.~\ref{virtual_hybrid}.

\end{document}